# Feasibility of 2D antiscatter grid concept for flat panel detectors: preliminary investigation of primary transmission properties

Cem Altunbas[1a], Yuncheng Zhong[2], Chris C. Shaw[2]

[1]Department of Radiation Oncology, University of Colorado School of Medicine, Aurora, CO, 80045

[2]Department of Imaging Physics, University of Texas M.D. Anderson Cancer Center, Houston, Texas 77030

[a] e-mail: caltunbas@gmail.com

**(This manuscript was written in 2014, but was not published previously)**

**Abstract:**

Suppressing the effects of scattered radiation in flat panel detector (FPD) based CBCT still remains to be a challenge. To address the scatter problem, we have been investigating the feasibility of a two dimensional antiscatter grid (2D ASG) concept for FPDs. Although a 2D ASG can potentially provide high scatter rejection capability, primary transmission characteristics of a 2D ASG and its implications in image quality plays a more critical role in implementation of the 2D ASG concept. Thus, in this work, a computational model was developed to investigate the primary transmission properties of the 2D ASG for various grid and FPD pixel geometries, and the improvement in signal to noise ratio (SNR) was calculated analytically to demonstrate the impact of 2D ASG's transmission characteristics on image quality.

Computational model showed that average primary transmission fraction ($T_P$) strongly depends on the septal thickness of 2D ASG, and 2D ASG can provide higher $T_P$ than existing radiographic ASGs at a septal thickness of 0.1 mm. Due to the higher $T_P$, 2D ASG was also predicted to provide SNR improvements in projections in low to moderate scatter environments typically observed in CBCT imaging. On the other hand, the model also indicated that the shadow or footprint of the 2D ASG leads to spatially nonuniform variations in primary signal in FPD pixels. Reduction of septal thickness and optimization of 2D ASG's pitch may play an essential role in reducing such variations in primary image signal, and avoiding potential image artifacts associated with 2D ASG's footprint.

## 1. Introduction

Scattered radiation is one of the major causes of image quality degradation in projection radiography and FPD based CBCT imaging [1,2]. To date, radiographic and fluoroscopic ASGs (referred as radiographic ASG in the rest of the text) and air gap methods remain to be the most commonly employed scatter rejection techniques. However, particularly in FPD based CBCT imaging, utilization of radiographic ASGs and air gaps provided only modest improvements in image quality. Improvements in CT number accuracy was not sufficient enough for quantitative imaging applications, such as radiation therapy dose calculations[3], and fell short of improving low contrast sensitivity. Radiographic ASGs were shown to reduce signal-to-noise ratio (SNR) and contrast-to-noise (CNR) when scatter intensity was moderate in projections (i.e. SPR ~ 1 or below) [4-7], which was attributed to low primary transmission through radiographic ASGs[8].

To better suppress the effects of scatter, we have been investigating the feasibility of using 2D ASGs in conjunction with FPDs. The proposed device consists of a 2D array of grid elements separated by radio-opaque septa. To account for x-ray divergence and maximize primary transmission, each grid element is aligned, or focused, towards the point x-ray source. The 2D ASG's septal thickness was envisioned to be in the order of 0.1 to 0.25 mm and the grid pitch was expected to be 2 to 5 mm. The grid ratio, i.e. the ratio of grid height to grid element width, is expected to be similar to the grid ratios of radiographic ASGs. We believe that the utility of 2D ASG is best suited for CBCT applications, as the focused 2D grid architecture requires fixed source-detector geometry for optimal primary transmission. Current amorphous silicon (aSi) FPDs do not utilize pixelated scintillators, akin to MDCT detectors, and hence, continuous layer of scintillator in a FPD decouples the 2D ASG's septa from the FPD's pixels. Therefore, shape (e.g. square, hexagonal) and pitch of grid holes may be determined independent from the FPD pixel geometry.

In recent years, advances in fabrication methods allowed manufacturing of 2D ASGs for mammography[9] and MDCT systems [10]. To the best of our knowledge, the literature on the characteristics of 2D ASGs do not exist besides the above mentioned studies. To assess the feasibility of 2D ASG concept in FPD based imaging applications, first, 2D ASGs' transmission characteristics need to be investigated. To achieve this aim, the primary objective of this work is to evaluate the primary transmission properties as a function of 2D ASG's geometry, and demonstrate its potential impact on SNR in projections. Understanding the primary transmission characteristics is particularly important in assessing the feasibility of 2D ASG concept; Although a 2D grid is expected to provide better scatter suppression capability than a 1D grid [10], poor primary transmission may adversely affect SNR in projections [8] and CNR in CBCT images [7]. Furthermore, as the proposed 2D ASG's septal thickness is relatively large (0.1 - 0.25 mm) in relation to FPD's pixel dimensions (0.2 – 0.4 mm), the image signal underneath the ASG's footprint would be reduced, and large spatial variations in primary signal would be introduced. In a worst case scenario, pixels that are fully obstructed by 2D ASG's septa would not receive sufficient signal and should be treated as dead pixels.

In thus study, we utilized a computational model to evaluate average primary transmission fraction and primary signal distribution in the FPD as a function of 2D ASG's grid pitch and septal thickness. Also, an analytical model for SNR was utilized to demonstrate the impact of 2D ASG's primary transmission properties on the projection image SNRs.

## 2. Materials and methods

### 2.1 Simulation of primary transmission in a computational model

Briefly, the aim of the computational model was to simulate projection images of a 2D ASG in contact with FPD by accounting for the primary x-ray attenuation through the 2D ASG and the effects of detector blurring. It was assumed that the 2D ASG's apertures were perfectly aligned, or focused, towards the x-ray focal spot, matching the divergence of primary x-rays. To model the physical footprint of 2D ASG, a template of a 2D hexagonal aperture array was generated with the desired grid pitch and septal thickness (Fig. 1(a)-(b)). This template served as the binary attenuation map of the primary beam, where pixel values were set to 100% within the apertures (i.e. 100% primary transmission) and to 0 underneath the septa (primary x-rays incident on the 2D ASG's septa were fully absorbed). To minimize the effects of finite pixel size on the accurate simulation of septal thickness, pixel dimensions of the 2D template was set significantly smaller than the septal thickness (e.g. Template pixel widths were 0.005 mm).

In the 2D template described above, the incident primary beam was assumed to be parallel to the focused septa. However, even for ideally focused grid elements, primary beam will not be parallel due to the divergence of the x-ray beam and finite size of the focal spot, and would lead to a small, but non-negligible magnification of septal thickness on the detector plane. As a result, the "apparent" septal thickness of the grid structure in projection images is larger than its physical septal thickness. To account for this effect, 2D ASG template at a given physical thickness ($d_{phys}$) was generated by using apparent septal thickness, $d_{app}$, such as the one shown in Fig. 1(b). $d_{app}$ was defined as,

$$d_{app} = d_{phys} + d_{eff} \qquad (1)$$

where $d_{eff}$, effective increase in septal thickness, is an empirically determined free parameter. $d_{eff}$ is also a function of the 2D ASG's height, but the effect of grid height was omitted in this first iteration of the computational model.

Blurring within the FPD was simulated in two steps. In the first step, the 2D ASG template was convolved by a point spread function, PSF, which accounted for the blurring due to the scintillator and detector glare. In the second step, blurring due to integration in detector pixels was achieved by binning the template's pixels to match the FPD's pixel size (Fig. 1(c)-(d)). For the PSF, the following function proposed by Poludniowski et al. [11] was employed,

$$PSF(r) = \frac{a_1}{2\pi b_1^2} e^{-r^2/2b_1^2} + \frac{a_2}{2\pi b_2^2} e^{-r/2b_2} + \frac{a_3}{2\pi b_3^2} \frac{1}{(1+r^2/b_3^2)^{3/2}} \quad (2)$$

where $a_i$ and $b_i$ are free parameters, and $r$ is radial distance from the symmetry center of the PSF. The PSF term accounted for blurring in the scintillator and detector backscatter, and excluded the effects of pixel size. The effects of focal spot associated blurring were implicitly included in the PSF term.

To determine the free parameters in the PSF and $d_{eff}$, a multi-hole collimator for gamma cameras was employed in lieu of a 2D ASG prototype, and its projection images were acquired by using a benchtop imaging system (see section 2.2 for more details). Free parameters of the model were determined by minimizing the difference between the cumulative pixel value histograms (PVH) of the simulated image ($PVH_{Model}(a_i, b_i, d_{add})$) and the acquired image of the multi-hole collimator ($PVH_{Measured}$) in the least squares sense,

$$minimize \| PVH_{Model}(a_i, b_i, d_{eff}) - PVH_{Measured} \|^2 \quad (3)$$

After the determination of free parameters, the computational model was used to simulate images of 2D ASGs with various grid pitches and septal thicknesses. For each 2D ASG geometry, average primary transmission fraction ($T_P$) values were calculated. $T_P$ was defined as,

$$T_P = \frac{I(with\ ASG)}{I(without\ ASG)} \times 100 \quad (4)$$

where *I* is the average of all pixel values within the simulated 2D ASG image. Without the 2D ASG in place, all pixels had a value of 100%. To evaluate the variation in primary image signal due to 2D ASG's footprint, PVH graphs were generated from each simulated image, such as the one shown in Fig. 3, and value of $PVH_{50}$ was calculated ( $PVH_{50}$: percentage of detector pixels that receive at least 50% of the flood field primary signal.). $PVH_{50}$ was chosen as a surrogate metric for the percentage of functional pixels (i.e. pixels that received less than 50% of flood field image signal were deemed dead pixels). The effects of 1X1 (pixel width: 0.194 mm) and 2X2 (pixel width: 0.388mm) pixel binning on the $PVH_{50}$ was also characterized using the same approach.

## 2.2 Image acquisition with the multi-hole collimator

To determine the free parameters in the computational 2D ASG model, projection image of a multi-hole collimator was utilized. The multi-hole collimator (Nuclear Fields Corp., Des Plaines, IL) (Fig. 2) was composed of abutting hexagon shaped grid apertures, or through-holes, parallel to each other, and they were separated by 0.25 mm thick lead septa. Each aperture was 2.5 mm in width and 32 mm in height, corresponding to a pitch and grid ratio of 2.75 mm and 12.8, respectively. Although, the parallel-hole collimator is not a substitute for 2D ASG due to the misalignment between the parallel holes and the divergent x-ray beam, its septal thickness, pitch, and grid ratio were similar to the envisioned 2D ASG. To avoid the misalignment problem, only a 1.2x1.2 cm² region encompassing the central axis of the beam was extracted from the projection image of the multi-hole collimator, where mismatch between parallel holes and beam divergence was negligible.

Images were acquired using a benchtop CBCT system optimized for surgical mastectomy specimen imaging [12, 13]. The system consisted of an aSi FPD with CsI scintillator (Paxscan 4030CB, Varian Medical Systems, Palo Alto, CA) and a radiographic x-ray tube with a focal spot size of 0.6 mm. Source to detector distance was 109 cm, and the irradiated field of view was fixed at 24 x 36 cm² at the detector plane. The multi-hole collimator was placed directly on the FPD's protective cover and it was aligned with central axis of the beam (i.e. the line that goes through the focal spot and detector plane, and orthogonal to the detector plane). 5 image frames were acquired with and without the 2D ASG in place at 120 kVp, flat field corrected [14], and averaged.

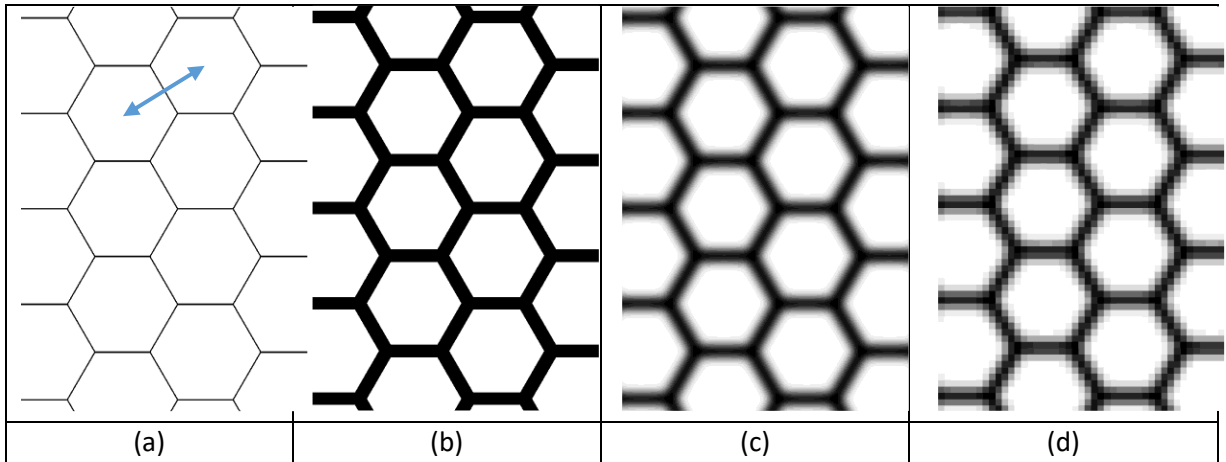

(a) (b) (c) (d)

**Figure 1.** The simulation of 2D ASG projection images in the 2D ASG model. First, a binary aperture array template was generated at the desired grid pitch (a). The pitch was defined as the distance between the centers of two adjacent apertures. In (b), the apparent septal thickness, $d_{app}$, in the template was adjusted to the desired value using a dilation operation. In (c), the template was blurred by convolving (b) with the detector PSF. In (d), template pixels were binned to match the detector pixel size and to simulate the blurring due to pixel aperture function.

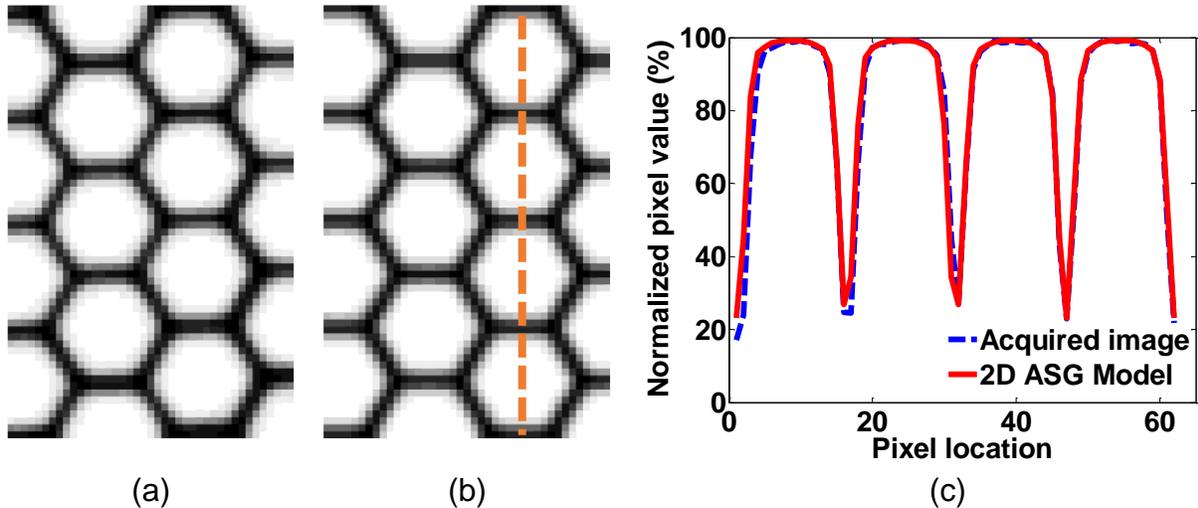

(a) (b) (c)

**Figure 2.** Acquired (a) and simulated (b) primary transmission images of the multi-hole collimator. The window/level was set identical for both (a) and (b). Pixel value profiles from the acquired and the simulated images (along the dotted line indicated in (b)) are shown in (c).

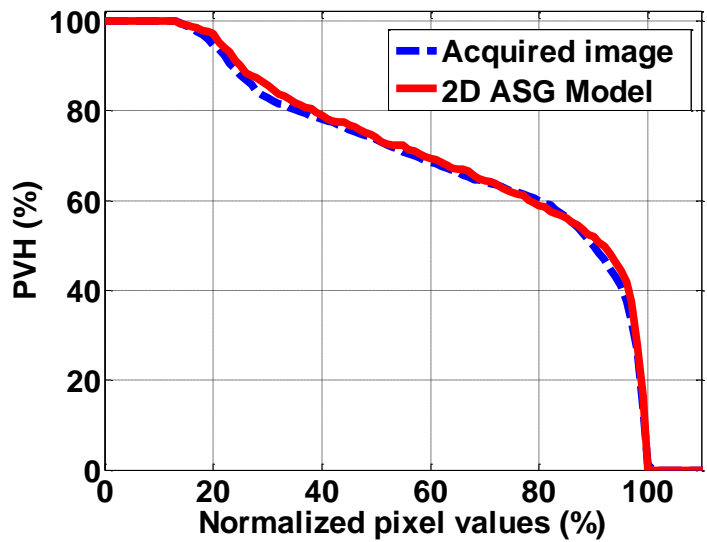

**Figure 3.** PVH graphs were generated from the acquired and simulated images of the multi-hole collimator (see Fig. 2).

### 2.3 The effect of 2D ASG on the projection image SNR

To demonstrate the impact of 2D ASG on image quality, SNR improvement factors, $K_{SNR}$, for a hypothetical radiographic ASG and 2D ASG were calculated by using the formalism described by Fetterly et al. [15]. $K_{SNR}$ is the ratio of SNRs with and without the ASG in place, and it is defined as,

$$K_{SNR} = \frac{(1+SPR)}{B^{0.5}(1+SPR_G)} \quad (5)$$

Bucky factor, B, and $SPR_G$ (scatter-to-primary ratio with ASG in place) are defined as

$$B = \frac{(1+SPR)}{T_P(1+SPR_G)} \quad (6)$$

$$SPR_G = \frac{T_S}{T_P} SPR \quad (7)$$

For a given ASG with known $T_P$ and $T_S$ values, $K_{SNR}$ was calculated as a function of SPR at the detector plane.

## 3. Results

### 3.1 Evaluation of primary transmission with the 2D ASG model

Free parameters of the PSF and $d_{eff}$ were determined (Table 1) using the method described in Section 2.1. Acquired and simulated images of the multi-hole collimators were shown in Fig. 2. Measured $T_P$ and $PVH_{50}$ were 72.8% and 73.4%, and simulated values of $T_P$ and $PVH_{50}$ were 73.6% and 74.3%, respectively. The agreement between the acquired and simulated images was also qualitatively evident in the primary signal profiles (Fig. 2(c)), and their PVH graphs (Fig. 3).

Table I. Free parameters of the $PSF$ and $d_{eff}$

| Parameter | Value (mm) |
|---|---|
| $a_1$ | 0.7769 |
| $a_2$ | 0.1781 |
| $a_3$ | 0.0450 |
| $b_1$ | 0.1365 |
| $b_2$ | 0.8138 |
| $b_3$ | 6.3889 |
| $d_{eff}$ | 0.1400 |

Primary transmission properties of 2D ASGs were evaluated for septal thicknesses ranging from 0.1 to 0.25 mm, and grid pitches between 2 and 5 mm. In Fig. 4, calculated $T_P$ values were indicated by markers, and lines represent polynomial fits to points. $T_P$ showed strong dependence on both septal thickness and grid pitch. For instance, $T_P$ was 65.3% at a grid pitch of 2 mm and a physical septal thickness of 0.25 mm. $T_P$ increased to 91.1% by reducing the grid pitch from 2 to 5 mm, and reducing the septal thickness from 0.25 to 0.1 mm.

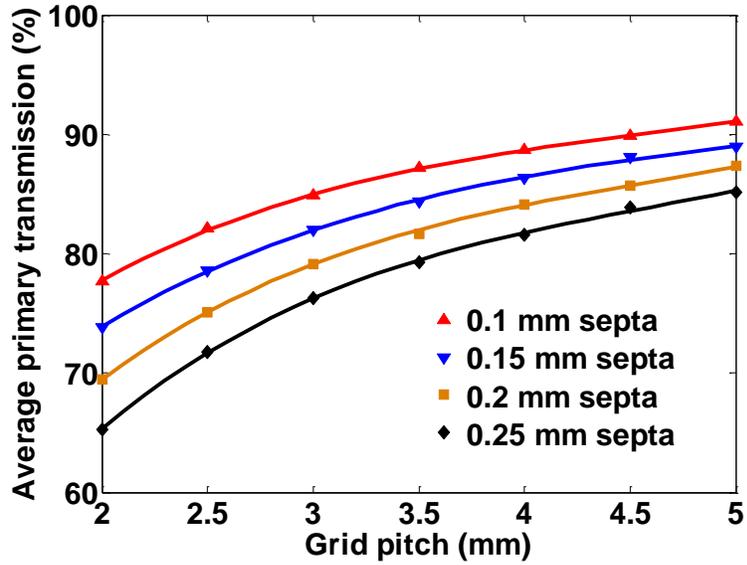

**Figure 4.** Average primary transmission fraction ($T_P$) was calculated as a function of septal thickness and grid pitch in the 2D ASG model.

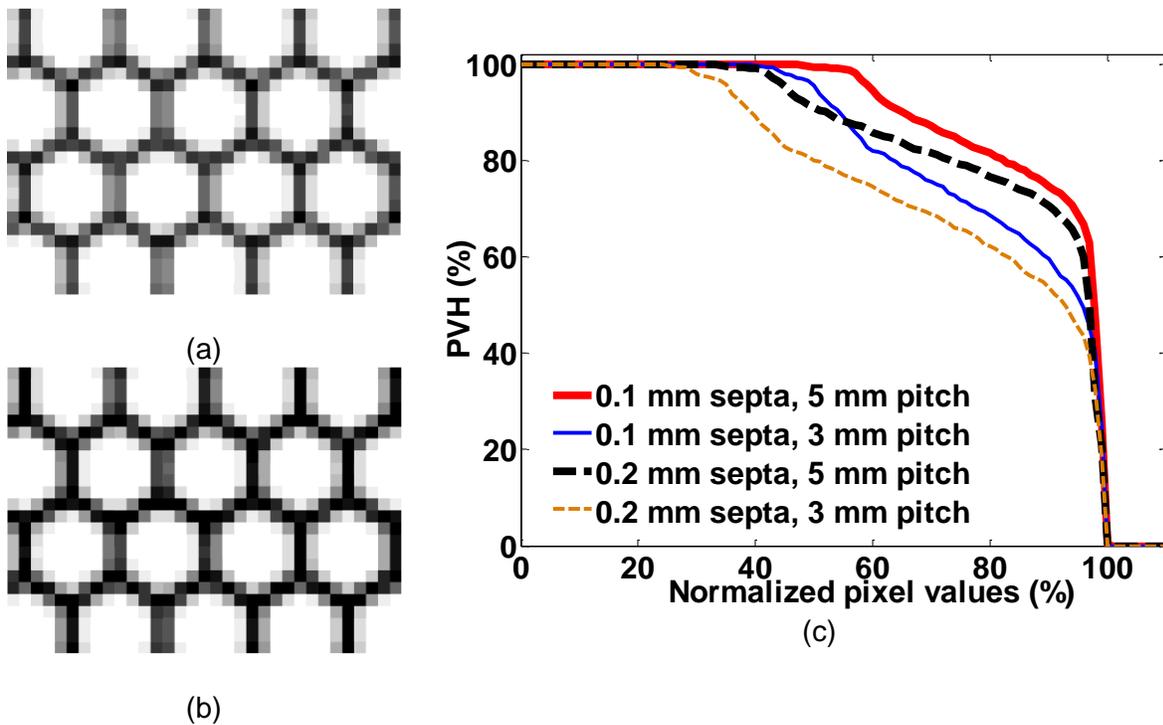

**Figure 5.** Simulated images of 2D ASGs with septal thicknesses of 0.1 and 0.2 mm are in (a) and (b), respectively. The grid pitch (3 mm) and the detector pixel size (0.388 mm) were the same in both images, and images are displayed at the same window/level. The PVHs of (a), (b) are shown in (c). PVH of 2D ASGs with a grid pitch of 5 mm are also shown in (c).

Spatially non-uniform reduction in primary signal due to the 2D ASG's footprint was demonstrated in Fig. 5; the 2D ASG models in Figs. 5(a) and 5(b) have septal thicknesses of 0.1 mm and 0.2 mm, respectively, have the same grid pitch (3 mm), and images have the same detector pixel width (0.388 mm). Both images are displayed at same window/level. At 0.2 mm septal thickness, relatively lower image signal underneath the footprint was evident in simulated images, and in their corresponding PVHs (Fig. 5(c)). By reducing the septal thickness from 0.2 to 0.1 mm, $PVH_{50}$ increased from 79.9% 95.6%. The reduction in primary signal was further minimized by reducing the grid pitch from 3 to 5 mm Fig. 5(c)).

Percentage of functional pixels (i.e. $PVH_{50}$) as a function of septal thickness, grid pitch, and detector pixel size are shown in Fig. 6. Septal thickness and the grid pitch had the most impact on $PVH_{50}$. For example, at a pixel size of 0.194 mm, $PVH_{50}$ increased from 68.1 to 88.7% by reducing the septal thickness 0.2 to 0.1 mm, and reducing the grid pitch from 2 to 3 mm.

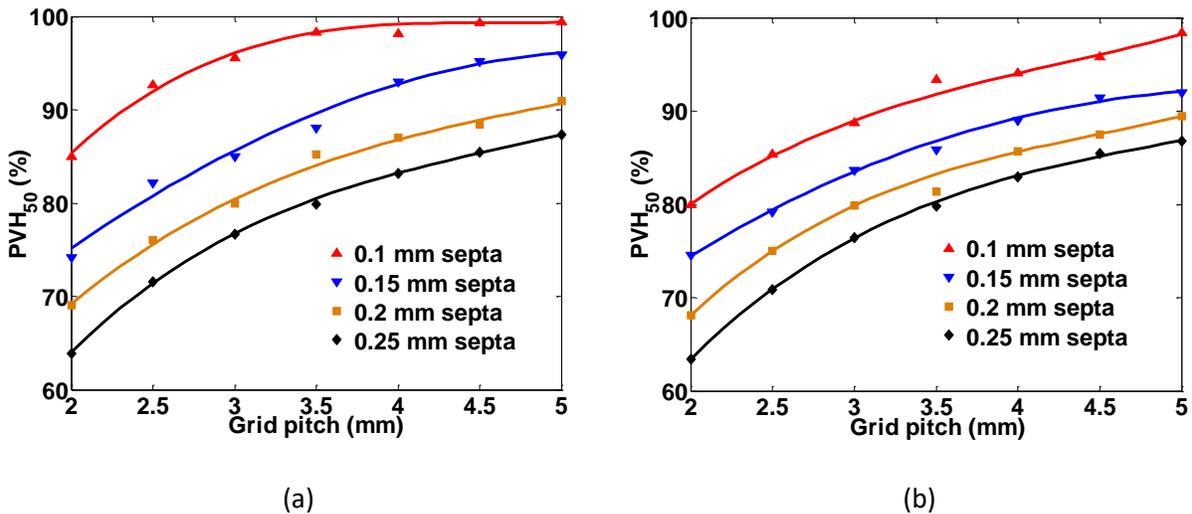

(a) (b)

**Figure 6**. $PVH_{50}$ were calculated as a function of grid pitch and septal thickness. Figures (a) and (b) correspond to FPD pixels widths of 0.194 and 0.388 mm, respectively. Markers indicate the calculated values by the model, and lines are cubic polynomial fits.

### 3.3 Improvement of SNR in projection images

$K_{SNR}$ for two 2D ASGs and a radiographic ASG were plotted as a function of SPR on the detector plane in Fig. 7. $K_{SNR} = 1$ was marked by the dotted horizontal line, indicating that SNR was improved by using an ASG above this line. Based on our results in Section 3.2, $T_P$ of both 2D ASGs were assumed to be 85%, and $T_S$ were 6% and 10%, respectively. Due to the absence of existing literature, $T_S$ values for 2D ASGs were conservatively estimated from the radiographic ASG literature. For the radiographic ASG, $T_P$ and $T_S$ were obtained from the reported values in the literature [15, 16], and they were 70% and 10%, respectively.

In Fig.7, at high SPR imaging conditions (e.g. SPR>1), both radiographic and 2D ASGs provided improvement in SNR, albeit $K_{SNR}$ provided by the radiographic ASG was lower. The advantage of 2D ASG was more apparent in low to moderate SPR conditions; $K_{SNR}$ of radiographic ASG was below 1 when SPR was less than 0.54. On the other hand, $K_{SNR}$ of both 2D ASGs remained above 1 at SPR levels down to 0.2. $T_S$ of 2D ASGs (10% vs. 6%) had minimal impact on $K_{SNR}$ at low SPR conditions. With respect to the radiographic ASG, 2D ASG with a $T_S$ of 10% provided $K_{SNR}$ improvements at lower SPR values, indicating that higher $T_P$ of 2D ASG played a more important role than $T_S$ in improving SNR, particularly in low SPR imaging conditions.

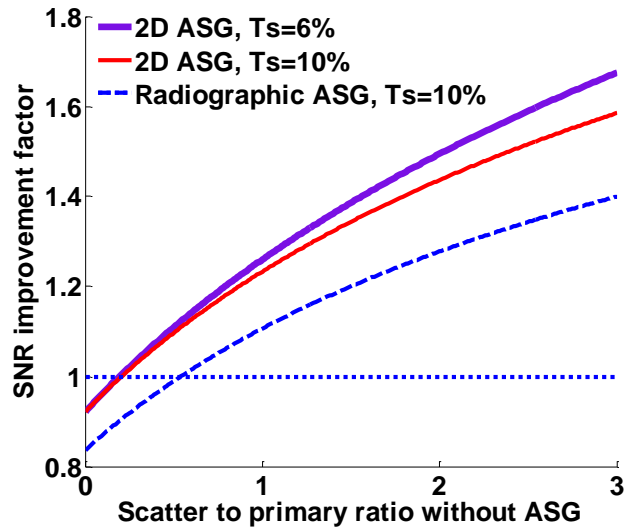

**Figure 7.** SNR improvement factors ($K_{SNR}$) for two hypothetical 2D ASGs and a radiographic ASG. A $K_{SNR}$ value larger than 1 indicates that SNR is improved with the use of ASG. $T_P$ of the 2D ASGs and the radiographic ASG were 85% and 70%, respectively.

## 4. Discussions

In this study, we presented a new 2D ASG concept for FPD based imaging applications, and performed proof of concept evaluations to demonstrate its primary transmission characteristics. 2D ASG's $T_P$ was expected to be in the order of 70-90% depending on the grid geometry, whereas reported $T_P$ value of radiographic ASGs in the literature was typically in the range of 50-75% depending on the grid type and imaging conditions.[4, 15-17]. As demonstrated in Section 3.2, higher $T_P$ provided by 2D ASG is an important factor in improving SNR in low to moderate SPR conditions (i.e. SPR is ~1 or less). In CBCT imaging, the SPR is around 1 in majority of imaging conditions [6], and therefore, 2D ASG may be favorable over radiographic ASGs: a 2D ASG may potentially improve SNR in projections and CNR in CBCT images [18].

A major challenge in the implementation of 2D ASGs in imaging systems lies in addressing the spatially nonuniform reduction in primary signal due to 2D ASG's footprint. Percentage of functional pixels (PVH$_{50}$) can be below 70% depending on the grid geometry, however PVH$_{50}$ can be increased to 95% or above, if the 2D ASG's septal thickness is reduced to 0.1 mm, the grid pitch is 3 mm or lower, and 2X2 pixel binning mode is used.

In this work, PVH$_{50}$ metric was introduced to characterize reduction in primary signal and to estimate percentage of functional detector pixels (i.e. pixels that are not treated as "dead" pixels), which was deemed to be a conservative metric. PVH$_{50}$ was not intended to be a "definitive" metric in determining the percentage of functional pixels; it was rather aimed to demonstrate the reduction in primary signal and its potential impact on the functional pixels in a simplified form. In a realistic imaging scenario, determination of functional pixels will likely to depend on the desired noise and spatial resolution properties of the projection or CBCT images, and whether the reduced image signal in pixels can be accurately corrected. Potentially, reduced image signal may be corrected by using flat field correction techniques [14]. However, in the context of CBCT imaging, correction of 2D ASG's footprint in the presence of gantry sag may not be fully achieved with this standard approach. In the presence of gantry sag, or flex, 2D ASG's footprint would project onto slightly different locations within the FPD, as the location of focal spot would shift in relation to the FPD during gantry rotation. As a result, image signal in pixels underneath the ASG's footprint would also vary as a function of gantry sag, and it may not be accurately corrected using standard flat field correction techniques. Suboptimal recovery of signal in such pixels may lead to image artifacts, such as ring artifacts, in CBCT images. Gantry sag is also an issue for radiographic ASGs in CBCT systems, and gantry angle-specific flat field corrections were implemented to mitigate this problem [19, 20]. In principle, this approach may also be applied to correct the 2D ASG's footprint in CBCT projections.

How to fabricate a 2D ASG with thin septa (~0.1 mm) is a question remains to be answered. "Divergent" multi-hole collimators for gamma cameras [21, 22] is physically similar to the proposed 2D ASG with focused grid elements. Lead casting technology for divergent gamma camera collimators may be adapted for fabrication of 2D ASGs. Based on the information in literature, grid with septal thickness of 0.23-0.25 mm can be fabricated with lead casting, however minimum achievable septal thickness with this technology has not been established. Alternatively, 2D ASGs for MDCT systems [10] were manufactured by using a lithographic molding process, where septal thicknesses down to 0.07 mm were achieved. This approach appears to be promising, however fabrication of the proposed 2D ASG is still yet to be demonstrated.

## 5. Conclusions

This study was the first step towards determining the feasibility of 2D ASG concept for FPD based imaging applications. Our work indicates that 2D ASG can provide higher average primary transmission fraction, which is expected to lead to higher SNR in projections. On the other hand, 2D ASG's footprint leads to spatial variations in primary signal due to its relatively large septal thickness and low pitch. If such primary signal variations are severe enough, dead FPD pixel zones underneath the 2D ASG's footprint may be introduced. Our study suggests that reduction of septal thickness and grid pitch, and increasing the FPD pixel size play a key role in minimizing this problem.

In the context of FPD based CBCT systems, potentially lower scatter transmission through 2D ASG may provide higher CT number accuracy. Moreover, higher primary transmission through 2D ASG is expected to improve SNR in projections, and in return, may improve CNR in CBCT images. This is a particularly important advantage of the 2D ASG, since improvements in CNR with existing scatter suppression methods appear to be limited in typical CBCT imaging conditions.

**Acknowledgments**

Authors would like to thank Nuclear Fields Corp. (Des Plaines, IL) for providing multi-hole collimator samples.**References**

1. Siewerdsen, J.H. and D.A. Jaffray, *Cone-beam computed tomography with a flat-panel imager: magnitude and effects of x-ray scatter.* Medical Physics, 2001. **28**(2): p. 220-31.

2. Ruhrnschopf, E.-P. and K. Klingenbeck, *A general framework and review of scatter correction methods in x-ray cone-beam computerized tomography. Part 1: Scatter compensation approaches.* Medical Physics, 2011. **38**(7): p. 4296-4311.

3. Hatton, J., B. McCurdy, and P.B. Greer, *Cone beam computerized tomography: the effect of calibration of the Hounsfield unit number to electron density on dose calculation accuracy for adaptive radiation therapy.* Physics in medicine and biology, 2009. **54**(15): p. N329.

4. Wiegert, J., M. Bertram, D. Schaefer, N. Conrads, J. Timmer, T. Aach, and G. Rose, *Performance of standard fluoroscopy antiscatter grids in flat-detector-based cone-beam CT.* Proc. SPIE, 2004. **5368**: p. 67-78.

5. Kwan, A.L., J.M. Boone, and N. Shah, *Evaluation of x-ray scatter properties in a dedicated cone-beam breast CT scanner.* Medical Physics, 2005. **32**(9): p. 2967-75.

6. Sisniega, A., W. Zbijewski, A. Badal, I.S. Kyprianou, J.W. Stayman, J.J. Vaquero, and J.H. Siewerdsen, *Monte Carlo study of the effects of system geometry and antiscatter grids on cone-beam CT scatter distributions.* Medical Physics, 2013. **40**(5): p. -.

7. Lazos, D. and J.F. Williamson, *Monte Carlo evaluation of scatter mitigation strategies in cone-beam CT.* Medical Physics, 2010. **37**(10): p. 5456-70.

8. Neitzel, U., *Grids or air gaps for scatter reduction in digital radiography: a model calculation.* Medical Physics, 1992. **19**(2): p. 475-81.

9. Makarova, O.V., N.A. Moldovan, C.-M. Tang, D.C. Mancini, R. Divan, V.N. Zyryanov, D.G. Ryding, J. Yaeger, and C. Liu, *Focused two-dimensional antiscatter grid for